# Template nanowires for spintronics applications: nanomagnet microwave resonators functioning in zero applied magnetic field

A. Mourachkine,[1,2,*] O. V. Yazyev[3], C. Ducati[4] and J.-Ph. Ansermet[5]

[1]University of Cambridge, J. J. Thomson Ave., Cambridge CB3 0HE, UK

[2]Alternative Energy Research Institute, Cambridge, UK

[3]ISIC, Ecole Polytechnique Fédérale de Lausanne, CH-1015, Lausanne, Switzerland

[4]Department of Materials Science and Metallurgy, University of Cambridge, Pembroke Street, Cambridge CB2 3QZ, UK

[5]IPN, Ecole Polytechnique Fédérale de Lausanne, CH-1015, Lausanne, Switzerland

*Corresponding author. E-mail: andrei_mourachkine@yahoo.co.uk

# Abstract

Low-cost spintronic devices functioning in zero applied magnetic field are required for bringing the idea of spin-based electronics into the real-world industrial applications. Here we present first microwave measurements performed on nanomagnet devices fabricated by *electrodeposition* inside porous membranes. In the paper, we discuss in details a microwave resonator consisting of three nanomagnets, which functions in *zero* external magnetic field. By applying a microwave signal at a particular frequency, the magnetization of the middle nanomagnet experiences the ferromagnetic resonance (FMR), and the device outputs a measurable direct current (spin-torque diode effect). Alternatively, the nanodevice can be used as a microwave oscillator functioning in zero field. In order to test the resonators at microwave frequencies, we developed a simple measurement set-up.

\*\*\*\*\*\*\*\*\*\*\*\*\*\*\*\*\*\*\*\*\*\*\*\*\*\*\*\*\*\*\*\*\*\*\*\*\*\*\*\*\*\*\*\*\*\*\*\*\*\*\*\*\*\*\*\*\*\*\*\*\*\*\*\*\*\*\*\*\*\*\*\*\*\*\*\*

**Table of Content figure**

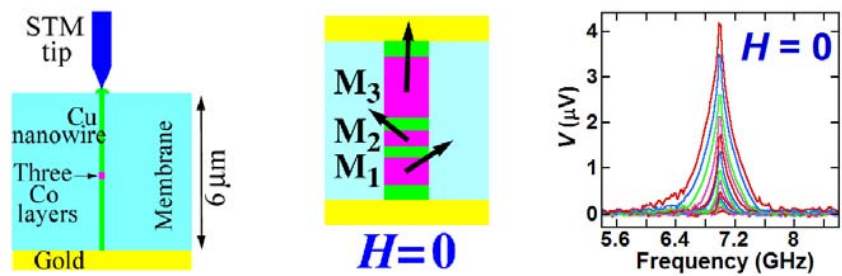

\*\*\*\*\*\*\*\*\*\*\*\*\*\*\*\*\*\*\*\*\*\*\*\*\*\*\*\*\*\*\*\*\*\*\*\*\*\*\*\*\*\*\*\*\*\*\*\*\*\*\*\*\*\*\*\*\*\*\*\*\*\*\*\*\*\*\*\*\*\*\*\*\*\*\*\*

MANUSCRIPT TEXT

Spintronics is the emerging technology of spin-based electronics[1]. It is anticipated that spintronics will offer new functionalities that go beyond those possible with charge-based electronic devices. *Low-cost* spintronic devices functioning in *zero* applied magnetic field are required for bringing the idea of spin-based electronics into the real-world industrial applications.

In addition to the phenomenon of the spin-transfer torque by spin polarized direct current (dc), predicted theoretically[2,3] and recently observed experimentally[4-6], the application of an alternating current (ac) at a particular frequency to a magnetic nanostructure (e.g. spin valve) can excite the ferromagnetic resonance (FMR) of one of the nanomagnets, leading to the generation of a measurable dc current (spin-torque diode effect)[5,6]. However, all these nanodevices[4-6] were tested in external magnetic field.

Recently, a nanodevice consisting of three ferromagnets was theoretically proposed as a unit of magnetic random access memory (MRAM) functioning in zero magnetic field[7]. This device was just tested to excite the steady state oscillations (SSO) of the magnetization of the middle nanomagnet[8,9]. The magnetization switching in the three-nanomagnet device has been recently measured too[10]. At the same time, no microwave measurements (spin-torque diode effect) in zero applied magnetic field, $H = 0$, have been reported yet. Here we present microwave measurements performed on three nanomagnet devices in zero magnetic field. To our knowledge, these are the first microwave measurements performed on nanodevices fabricated by electrodeposition inside porous membranes[11]. In order to test the three nanomagnets resonators at microwave frequencies, we developed a simple measurement set-up.

For practical applications, the advantage of the use of the structure with three magnetic layers in comparison with a standard two-layer spin valve is twofold: (i) the presence of external magnetic field is not necessary—it can function in *zero* magnetic field, and (ii) the magnitude of the dc current-driven spin torque is enhanced.

We examine samples made by electrochemically depositing a multilayer of composition 25 nm $Co_{85}Cu_{15}$/ 10 nm Cu/ 7 nm $Co_{85}Cu_{15}$/ 10 nm Cu/ 100 nm $Co_{85}Cu_{15}$ as depicted in Figure 1b. To fabricate template nanowires, we used the single-bath method. A review on the fabrication of template nanowires can be found elsewhere[12]. In short, a solution of $CuSO_4·5H_2O$ (0.0107 mol)/$CoSO_4·7H_2O$ (0.427 mol)/$H_3BO_3$ (0.728 mol) with a pH of about 3.3 is used at 35°C. For Co and Cu deposition, a potential of -1V and -0.4V is applied, correspondently. The typical deposition rates are 260 nm/s for Co and 25 nm/s for Cu. In our nanodevices, the magnetizations of the ferromagnets are not fixed (e.g. by exchange baising), therefore, we shall call our three-ferromagnet spintronic device as a sandwiched *pseudo*-spin-valve (PSV). Figure 1a sketches the equilibrium orientations of the magnetizations of a sandwiched PSV in *H* = 0. The multilayer structures are grown in the middle of membrane pores having a diameter of 40±5 nm (ref. 13, also see Supporting Information), and electrically connected by Cu nanowires, as sketched in Figure 1c. We utilized commercially-available porous polycarbonate ion-track etched membranes (Osmonics Inc.) having a thickness of 6 μm. The pore density is about $6 \times 10^6$ pores/cm$^2$. The single-bath method yields the presence of about 15% of Cu in Co layers[14]. $Co_{85}Cu_{15}$ has the fcc structure[14] which is an advantage because its magnetocrystalline anisotropy is smaller than that of the hcp Co. Furthermore, the saturation magnetization of $Co_{85}Cu_{15}$ is significantly smaller than that of bulk Co (~16 kOe)[14]. To our knowledge, our devices have the smallest cross-section, ø40 nm, in comparison with others presented in the literature, exhibiting the FMR or SSO.

It is worth noting that electrodeposition has a number of advantages to fabricate spintronic devices: it is a simple, low-cost technique which does not require clean rooms, sputtering installations and any post-processing[11,12]. One sample is made at ambient conditions in less than 10 minutes. By using this technique, one can fabricate spintronic devices, literally, on a kitchen table.

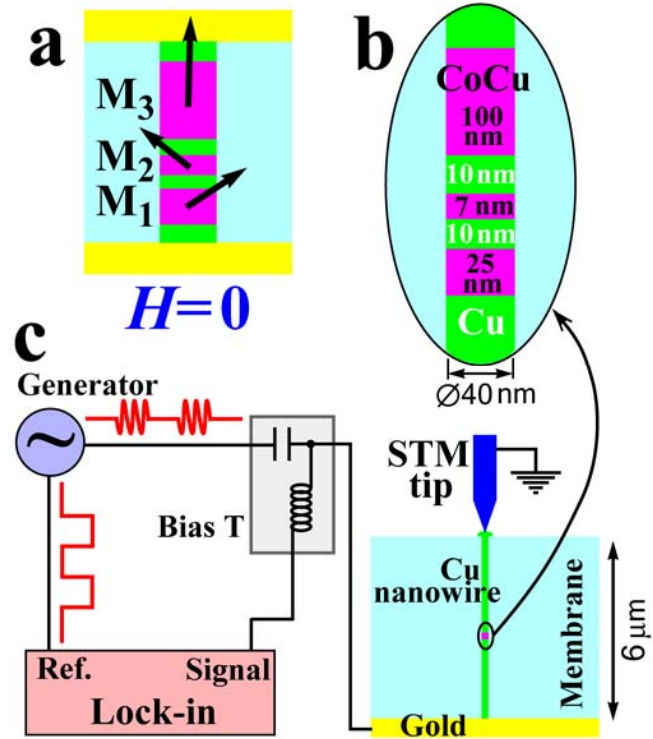

**Figure 1.** (a) A pseudo-spin-valve ($M_1$ and $M_2$ nanomagnets) with an added cylinder ($M_3$). The black arrows show the equilibrium orientations of the magnetic moments $M_1$, $M_2$ and $M_3$ for each nanomagnet in zero applied field. (b) A part of the cross-sectional view of a nanowire sketched in panel (c). The nanowire is electrodeposited in a pore of a polycarbonate ion-track etched membrane with a thickness of 6 μm. The spintronic device (25 nm $Co_{85}Cu_{15}$/ 10 nm Cu/ 7 nm $Co_{85}Cu_{15}$/ 10 nm Cu/ 100 nm $Co_{85}Cu_{15}$) is embedded into Cu nanowires. (c) Schematic of circuit used to measure the FMR (ref. 6). The sample is inserted into a standard SMA microwave connector (see Figure 2). The nanowire is connected to the set-up by a golden layer sputtered on one side of the membrane and a STM tip at the other side (STM = surface tunneling microscope). A differential screw is used to approach the tip to the hemispherical top of a nanowire. The generator frequency is $2 \leq f \leq 18$ GHz. The chopping frequency is 1.5 kHz.

\* \* \*

Electrical contacts to the nanowires grown in porous membranes are conventionally made by two gold layers sputtered at the both sides of the membranes[15]. For high frequency applications, this method is inappropriate, since the two gold layers create a shunt capacitance. So, we had to develop a new technique to contact the nanowires to the measurement set-up. A gold layer was sputtered only at one side of the membranes, and a STM tip was utilized to connect the nanowires, one by one, at the other side of the membranes, as sketched in Figure 1c. To reduce the noise and to improve the impedance matching, each membrane with the spintronic devices was placed *inside* of a SMA microwave connector, as shown in Figure 2. By changing the direction of the tip approach, such a technique allows one to probe many of the millions of spintronic structures in one sample up to 40 GHz. We have tested 19 samples. All measurements have been done at room temperature.

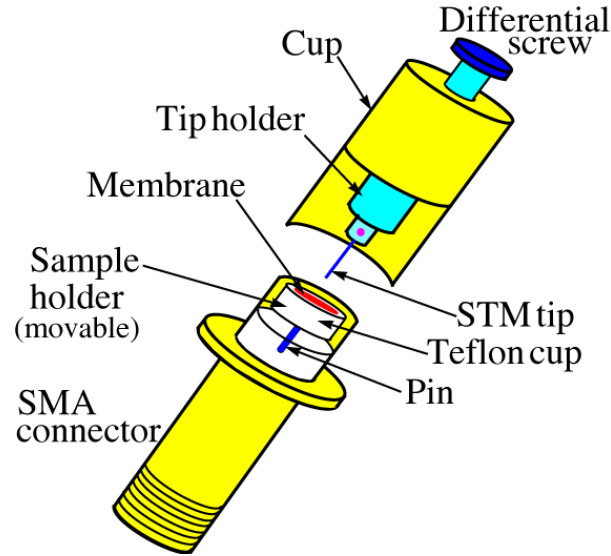

**Figure 2.** Sketch of in-house built test set-up to probe nanowires in porous membranes at high frequencies. The top of the SMA connector and the end of the cup are shown openly. A detailed description of the set-up can be found in Supporting Information.

\* \* \*

Figure 3a depicts the dc voltage response measured by passing 320 μA of ac current through a sandwiched PSV in zero magnetic field. The full width at half maximum (FWHM) is about 300 MHz which is more or less typical for our samples. The shape of the resonance curve is almost Lorentzian. However, we observed different shapes of the curves: in many cases, the shape of the resonance curve was not Lorentzian but rather asymmetric, "triangular", similar to that observed elsewhere[6] (a property of nonlinear oscillators). In our measurements, the resonance frequency of sandwiched PSV in $H = 0$, $f_{FMR}$, varied from sample to sample, and was in the interval $5 \leq f_{FMR} \leq 7$ GHz. Numerical simulations show that the variation of $f_{FMR}$ is related to the tiny variation of the diameter of pores: the larger the diameter, the higher the resonance frequency.

The magnetoresistances of this sandwiched PSV are depicted in Figure 3b, showing a small hysteresis at high |$H$|. In the general case, a magnetic structure showing a hysteresis has two remnant states. The inset in Figure 3a depicts two resonant curves measured at $H = 0$ after the sandwiched PSV was exposed to 8 kOe (red curve, as that in the main plot) and to –8 kOe (blue). The two curves in the inset of Figure 3a are very similar, which is an indication that the structure of this sandwiched PSV does not have major defects (otherwise, the curve shapes are different).

In order to understand better the behavior of the sandwiched PSV, we tested the sandwiched PSV in an external magnetic field applied perpendicular and parallel to the nanowires (thus, to $I_{rf}$). We start with the case $H \perp I_{rf}$. The evolution of the FMR spectrum depicted in Figure 3a in $H \perp I_{rf}$ is shown in Figure 3c ($H \leq 0$) and Figure 3d ($H \geq 0$). In Figures 3c and 3d, one can see that, in addition to the change of FMR frequency, at high |$H$| there appear a number of smaller features which will be discussed below. The FWHM of the main mode as a function of $H$ remains around 300 MHz, while the magnitude of the main mode has a maximum, 39 m$\Omega$/mA, at $H = 375$ Oe and decreases in both directions from 375 Oe.

To explain jumps in the magnetoresistances in Figure 3b, we performed numerical simulation of dipolar interactions among three nanomagnets in a sandwiched PSV in $H \perp I_{rf}$. The calculations of the angle between each magnetization and the nanowire as a function of $H$ are presented in Figure 4b (for details, see Supporting Information). The sketches of the orientations of each magnetization along the magnetoresistance are shown in Figure 3e. According to the simulations, by changing the field magnitude, $M_3$ simply rotates from +90º to –90º away from $M_1$ and $M_2$, and is responsible for the jump at -3.5 kOe. The other jumps are caused by sudden re-orientations of either $M_1$ (at 1.33 kOe and at –1.41 kOe) or both $M_1$ and $M_2$ simultaneously (at –0.76 kOe). Below -3.5 kOe, all nanomagnets are orientated along the field. At $H = 0$, the magnetization of $M_3$ lies almost along the nanowire; $M_1$ and $M_2$ have about -73º and 75º with the nanowire axis, respectively. Thus, the relative angle between $M_1$ and $M_2$ is about 148º. In Figure 3e, the red stripe shows the interval of the existence of the main FMR mode, $-0.62 \leq H \leq 1.4$ kOe. The "high-frequency mode" manifests itself exclusively on the small "plateau" at $-1.5 \leq H \leq -0.9$ kOe (cyan stripe).

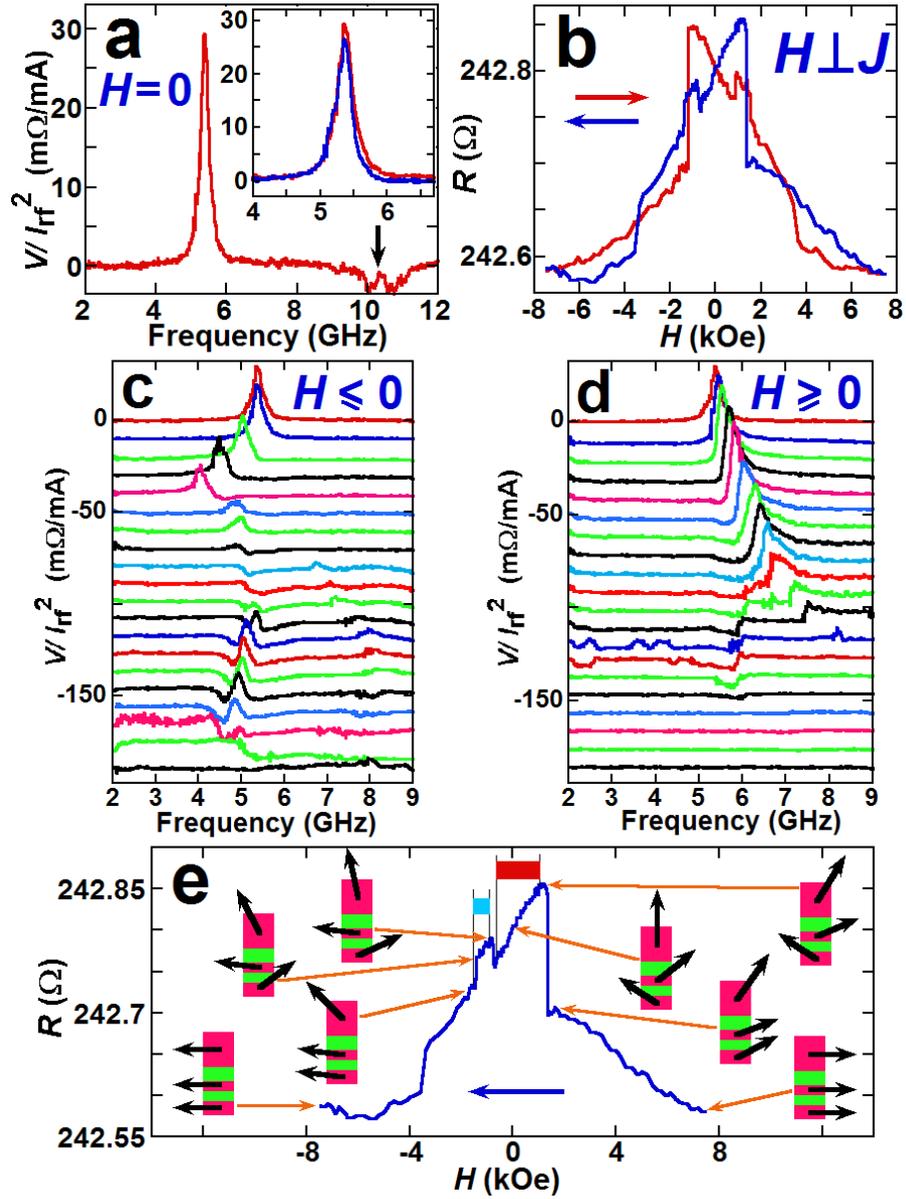

**Figure 3.** (a) The dc voltage generated by a sandwiched PSV in zero applied magnetic field as a function of frequency. The signal is normalized by the squared rms amplitude of applied radio frequency (rf) current, $I_{rf}$ = 320 µA. The FWHM is around 300 MHz. The spectrum obtained in applied magnetic field of $H$ = 8 kOe is taken as background and subtracted from the data. The spectrum is measured after the sandwiched PSV was exposed to $H$ = +8 kOe. The arrow indicates a small feature in the spectrum at 10.4 GHz. Inset in panel (a): the red curve as in the main panel, and the blue curve is the spectrum measured at $H$ = 0 after the sandwiched PSV was exposed to $H$ = -8 kOe. The labels of the axes correspond to the labels of the main panel. The data in panels (b), (c), (d) and (e) are obtained in the same sandwiched PSV. (b) Magnetoresistance curves of the sandwiched PSV. The magnetic field is applied perpendicular to the nanowire axis (thus, to $I_{rf}$). The arrows indicate the sweeping direction of the magnetic field in a loop, starting at –8 kOe. (c) Evolution of the spectrum depicted in panel (a) in applied magnetic field ($H \leq 0$). The field is perpendicular to the nanowire axis. The magnitudes of $H$ are

0 (top); -0.135; -0.25; -0.5; -0.62; -0.7; -0.75; -0.8; -0.9; -1; -1.12; -1.25; -1.375; -1.425; -1.5; -1.64; -1.7; -1.75; -1.875 and -2.5 kOe (bottom). Each spectrum is shifted by − 9.8 from the previous one. (d) As in panel (c) but at positive *H*. The magnitudes of *H* are 0 (top); 0.25; 0.375; 0.5; 0.6; 0.71; 0.8; 0.9; 1; 1.1; 1.2; 1.3; 1.4; 1.5; 1.6; 1.7; 1.8; 1.9; 2 and 3 kOe (botom). (e) The magnetoresistance curve from panel (b) and sketches of the orientations of the magnetic moments of each nanomagnet along the magnetoresistance curve, based on computer simulations (see Figure 4b). The field is applied perpendicular to the nanowire axis. The field sweep starts at 8 kOe. The red stripe shows the interval of the existence of the main FMR mode, -0.62 ≤ *H* ≤ 1.4 kOe (red curve in Figure 4a). The cyan stripe indicates the interval of the existence of the "high-frequency" FMR mode, -1.5 ≤ *H* ≤ -0.9 kOe (cyan curve in Figure 4a).

\* \* \*

The $f_{FMR}(H)$ dependences of the main mode and other smaller features in Figures 3c and 3d are depicted in Figure 4a. To explain the dependences, we computed solutions of the Landau-Lifshitz-Gilbert equation of motion for a single-domain magnet in the case of a small-amplitude ac current. First, we estimate the value of $4\pi M_{CoCu}$ (CoCu = $Co_{85}Cu_{15}$) as follows. We attribute a small feature in the spectrum in Figure 3a at 10.4 GHz to the FMR of $M_3$, in accordance with the numerical calculations (see Supporting Information). We use the Kittel equation for a finite-length cylinder with *H* applied along the axis of symmetry[16]:

$$f = \frac{g\mu_B}{h}(H + (N_x - N_z)M_{CoCu})$$

where $\mu_B$ is the Bohr magneton; *h* is the Planck constant; $N_x$ (= $N_y$) and $N_z$ are the demagnetizations. For a cylinder with 36×100 nm², $N_x - N_z$ = 0.29 (in units of 4π) (ref. 17). We use *g* = 2.25 of pure Co (ref. 16). At *H* = 0, we obtain $4\pi M_{CoCu}$ = 11.2 kOe. In order to find the FMR modes for each nanomagnet in $H \perp I_{rf}$, we search the local minima of the free energy for the system as a function of relative angles. The plot of the calculations matching the plot limits in Figure 4a is shown in Figure 4c (an extended picture is given in Supporting Information). The numerical calculations reproduce well the main FMR mode and the "high-frequency mode" in Figure 4a, caused by the FMR of $M_2$. The data in blue in Figure 4a, corresponding to a small feature with *negative* amplitude[6], are a manifestation of the FMR of $M_1$. In Figure 4a, the data in green reflecting small features in Figures 3c and 3d are not accounted for in Figure 4c. They may be manifestations of the week FMR resonance of $M_1$ (see Supporting Information).

Most likely, they reflect spatially *non-uniform* spin-wave modes of $M_3$ (Ref. 18) (Figure 4c depicts only *uniform* FMR modes).

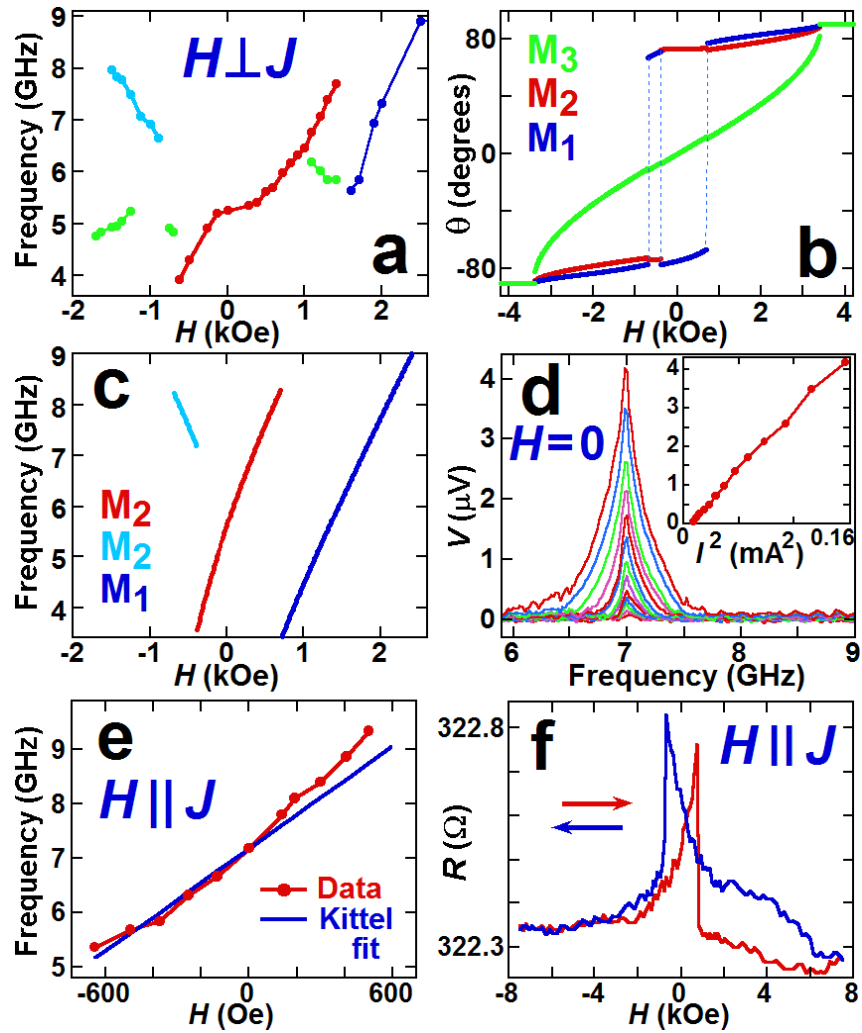

**Figure 4.** (a) Magnetic-field dependence of the main FMR mode (red curve) and other small features in the spectra in Figures 3c and 3d. The field is applied perpendicular to the nanowire axis. (b) Computer simulations for the sandwiched PSV sketched in Figure 1b: the angle between the magnetic moment of each nanomagnet and the nanowire axis as a function of magnetic field applied perpendicular to the nanowire axis. The data in blue, red and green indicate the $M_1$, $M_2$ and $M_3$ nanomagnets, correspondingly. At $|H| > 3.5$ kOe, the data overlap. The dashed lines indicate the jumps of $M_2$. (c) Computer simulations of FMR frequencies as a function of applied magnetic field for the sandwiched PSV sketched in Figure 1b. The field is applied perpendicular to the nanowire axis. The red and cyan curves correspond to the FMR of $M_2$ ("free layer"). The blue curve reflects the FMR of $M_1$. An extended picture of the FMR frequencies is given in Supporting Information. (d) dc voltage generated

by another sandwiched PSV in zero applied magnetic field versus frequency as a function of applied rf current. The difference in the microwave power supplied to the sample between two adjacent curves is −1 dBm, starting at −8 dBm from the top. The inset shows the linear dependence of the dc voltage on $I_{rf}^2$. The label of the vertical axis is identical to that of the main panel. (e) Magnetic-field dependence of the main FMR mode (red curve) and the linear Kittel-equation fit (blue line)[16] for another sandwiched PSV. The field is applied parallel to the nanowire axis. The data in panels (e) and (f) obtained in the same sandwiched PSV. (f) Magnetoresistances of the sandwiched PSV. The magnetic field is applied parallel to the nanowire axis. The arrows indicate the sweeping direction of the magnetic field in a loop, starting at −8 kOe.

<p style="text-align:center">*   *   *</p>

In Figure 4d, we show that the magnitude of dc voltage generated by $M_2$ is in accordance with the theoretical calculations[5,6], $V_{dc} \sim I_{rf}^2$. The dependence of FWHM of the curves on magnitude of $I_{rf}$ (or supplied microwave power) in Figure 4d is typical too[19]: at low $I_{rf}$ (the first eight curves from the bottom), the FWHM is practically $I_{rf}$-independent (= 110-140 MHz), while as $I_{rf}$ increases and becomes "high" enough, the FWHM starts to grow too (= 180-300 MHz). By increasing $I_{rf}$ further, we observed that the FMR frequency at a certain magnitude of $I_{rf}$ jumps to a higher frequency, and the shape of the resonance curve becomes asymmetric ("triangular")[19].

We also tested the sandwiched PSVs in $H \parallel I_{rf}$. In this case, the $f_{FMR}(H)$ dependence is in good agreement with the Kittel equation $f = g\mu_B(H - 4\pi M_{CoCu})/h$ (ref. 16), as shown in Figure 4e. The corresponding magnetoresistances are depicted in Figure 4f. For the $H \parallel I_{rf}$ case, the simulations of the dipolar interactions show that the jumps in the magnetoresistances at ±0.7 kOe are caused by a flip of $M_3$.

We would like to mention that the three-nanomagnet device sketched in Figure 1a was developed by one of us. After the experimental work was finished, we became aware of Ref. 9, and as a consequence, of Ref. 7.

To summarize, we presented microwave measurements performed on the three-nanomagnet spintronic device fabricated by electrodeposition inside porous membranes. We observed ac-current excited FMR of the middle nanomagnet in *zero* applied magnetic field. The resonance frequency of the nanostructure can be adjusted by external magnetic field, and its dependence on $H$ is in good agreement with numerical simulations. The simulations also allowed us to infer the behavior of the magnetizations of each nanomagnet in applied magnetic field. The magnitude of the dc voltage generated by the three-nanomagnet device is in good agreement with the theoretical predictions, $V_{dc} \sim I_{rf}^2$. In order to test the resonators at microwave frequencies, we developed a simple measurement set-up. The three-nanomagnet device can be used as a microwave detector, a microwave oscillator (see Supporting Information) or a unit of MRAM functioning in *zero* applied magnetic field. At room temperature, we were able to detect the FMR with an excellent signal-to-noise ratio of about $10^6$ Bohr magnetons only.

**Acknowledgement.** We thank M. Kläui, S. A. Wolf, I. N. Krivorotov, Y. Tserkovnyak and A. A. Kovalev for comments, and J. Van der Klink and A. Srivastava for help with measurements

**Supporting Information Available**. Experimental section, computer simulations, and additional measurements performed on sandwiched PSV, namely, the SSO excited by dc current.

# Supporting Information

## 1. Experimental Section

The samples are grown in a membrane placed beforehand on a metal platform with a diameter of 2.5 mm having a pin at the other side, as shown in Figure 2 of the main text and in Figure S1 below. The membrane is held on the platform by a Teflon cup having a hole with a diameter of about 1.6 mm. The nanowires are grown in the central part of the membrane. After the growth and drying procedure, the sample holder with the membrane is inserted into a standard microwave SMA female connector which is then closed by a metal cup (see Figure 2). The cup has a differential screw. The screw supported by a spring (not shown in Figure 2) pushes against a STM tip (Pt-Ir mechanically-sharpened wire). The tip is held by a tiny screw on the tip holder. The tip holder has "wings" which do not allow it to rotate, and make electrical contacts for the ac current. Once the SMA connector with the sample is closed by the cup, one can manually use the differential screw to approach the tip while watching a small-amplitude dc current to pass. The vertical resolution of the tip approach is approximately 7 μm. Since the density of nanowires is high, the tip often, but not always, lands on the hemispherical top of Cu nanowires, which has ø ~ 200-400 nm. The hemispherical top (end) of Cu nanowires is naturally formed at the membrane surface when the pores are completely filled with Cu. By changing the direction of the tip approach (by bending slightly the Pt-Ir wire or by turning the sample holder), such a technique allows one to probe many of the millions of spintronic structures in one sample up to 40 GHz.

**(a)**  **(b)**

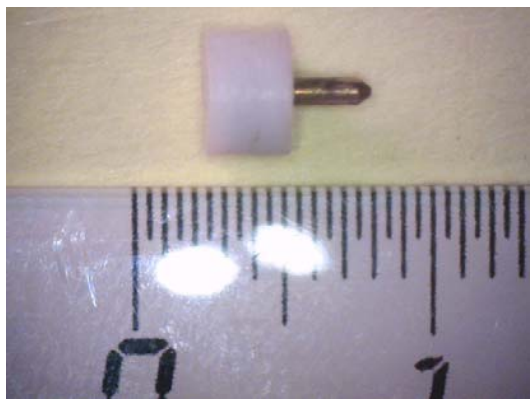 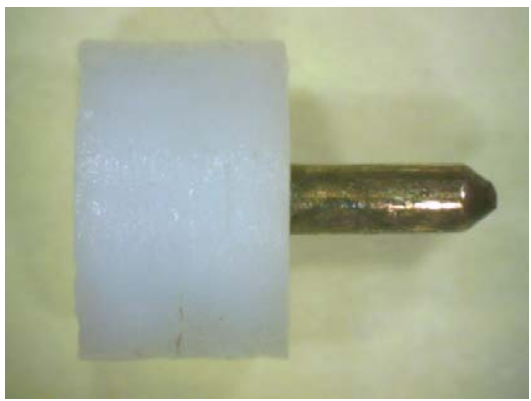

**(c)**  **(d)**

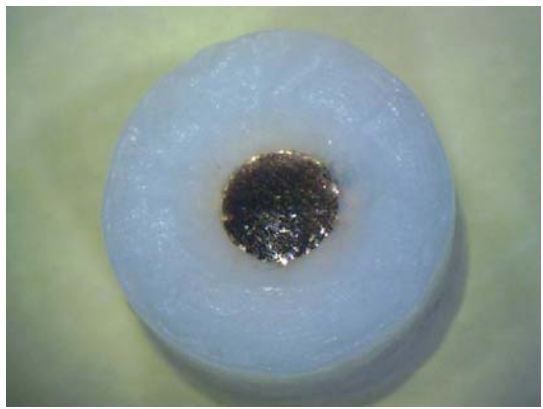 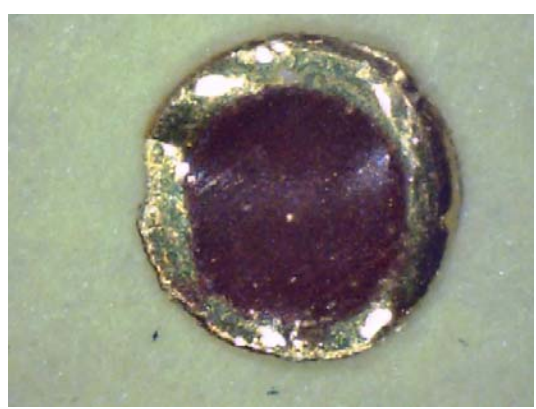

Figure S1. Sample holder (see Figure 2) and a polycarbonate membrane with nanowires. (a) and (b) Side view of a sample (membrane) holder. The scale in (a) is in centimeters. (c) Front view of a sample holder with a membrane. (d) Membrane with nanowires and a gold layer at the back side. The membrane thickness is 6 μm. The brown circle in the middle of the membrane is millions of electrodeposited Cu nanowires containing Co layers.

Figure S2 shows Cu nanowires with three Co layers. To picture the nanowires, a membrane with the nanowires used in the measurements was dissolved in chloroform, and the nanowires were deposited on holey carbon TEM grids. The nanowires were analysed by using a FEI Tecnai F20 microscope (200 kV acceleration voltage). Energy filtered TEM (EFTEM) was used to identify the Co regions within the Cu nanowires. Elemental (ratio) maps were acquired for cobalt first, by selecting the Co $M_{2,3}$ edge at 59.5 eV with a 10 eV energy slit (do not confuse $M_{2,3}$ with magnetizations $M_i$). The copper map of the same area was then acquired, at the Cu $M_{2,3}$ edge at 73.6 eV. The inner Cu/Co/Cu layers of the structure are very thin, so for a TEM, it is difficult to distinguish the inner layers. In addition, the Co layers contain about 15% of Cu.

(a)

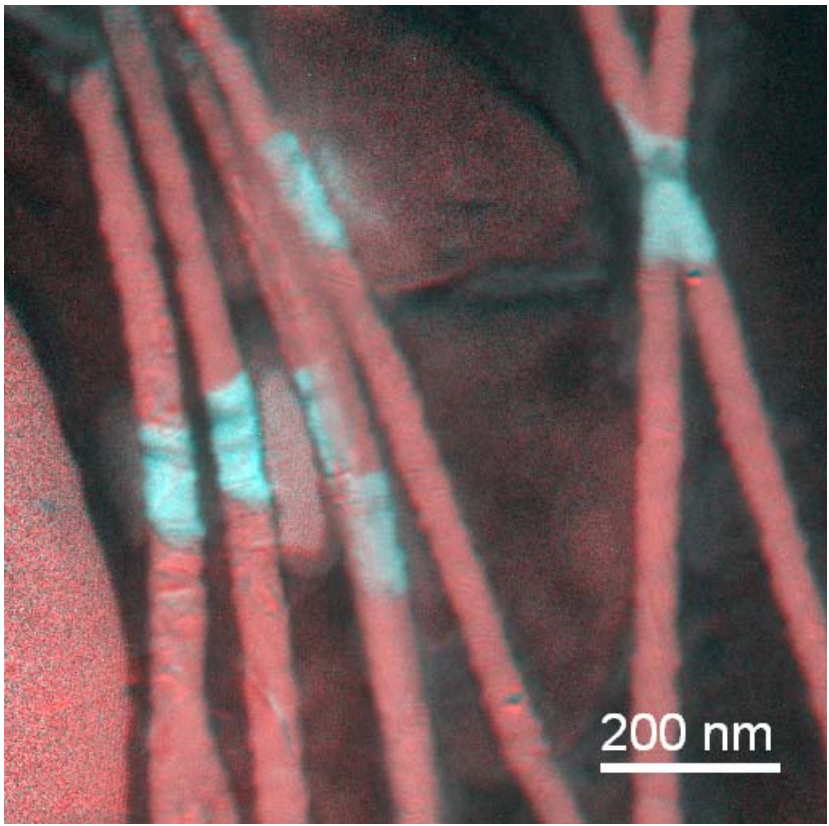

(b)

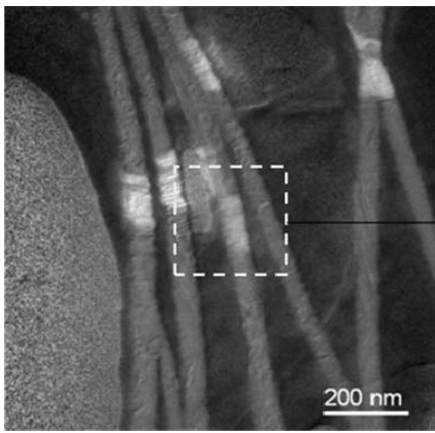

(c)

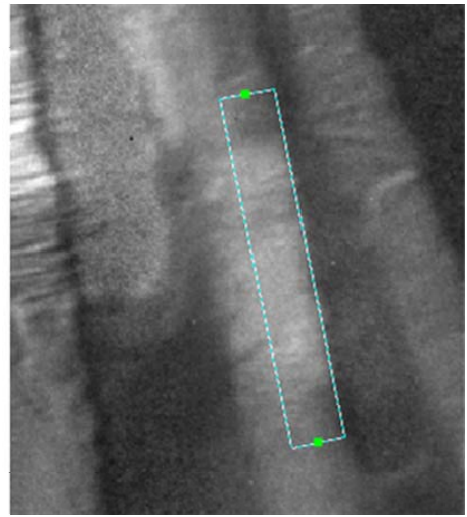

(d)

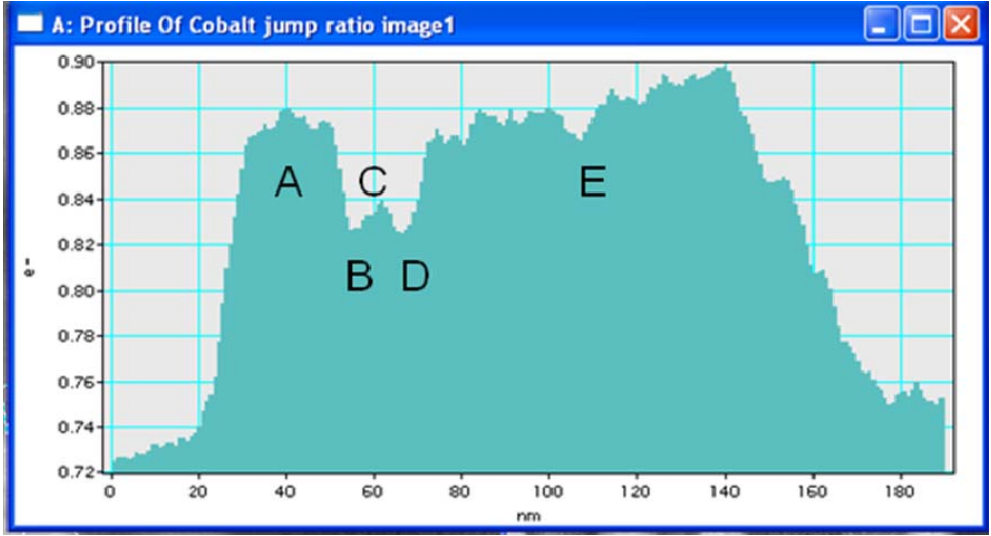

Figure S2. (a) TEM image of Cu nanowires with embedded Co layers (false-color energy filtered TEM (EFTEM) combined elemental maps for Cu and Co). In the image, some nanowires are broken while dissolving the membrane, some nanowires overlap. The cyan parts of the nanowires represent three Co layers. (b) Image in (a) in black and white. (c) Enlarged part of the image in (b). (d) The Co content along the marked path in (c). The widths of the Co layers are estimated from the profile of Co content, and about A ~ 25 nm; C ~ 7 nm, and E ~ 95 nm. The widths of the Cu interlayers are around B ~ 8 nm and D ~ 9 nm.

## 2. Numerical Simulations

To explain jumps in the magnetoresistances shown in the main text in Figure 3, and the $f_{FMR}(H)$ dependences depicted in Figure 4a for sandwiched PSVs, we performed numerical calculations.

The Landau-Lifshitz-Gilbert (LLG) equation of motion for a single-domain nanomagnet can be expressed in terms of the total free energy $F$ instead of effective fields. When the easy-axis directions of the various anisotropies or nanomagnets have different orientations, as in our case, the LLG computations with the energy terms are more straightforward. In our sandwiched PSV, the easy axis of the $M_3$ cylinder is perpendicular to the easy axes of the $M_1$ and $M_2$ nanomagnets.

The magnetic configurations of the system were found by searching all local minima of the total free energy: $F = F_H + F_D + F_S$, with respect to the magnetic-moment orientations of the three nanomagnets. The first term $F_H$ is the energy of the nanomagnets in applied magnetic field. $F_D$ is the energy of the dipolar interactions among the nanomagnets, and $F_S$ is the contribution of the shape anisotropy. Since $Co_{85}Cu_{15}$ in the fcc structure has a small magnetocrystalline anisotropy, in our calculations we neglect possible contributions of magnetocrystalline and interface anisotropies.

The effective dipolar interactions were determined by integrating numerically over a grid the magnetostatic interactions among the nanomagnets. The found effective distances of the dipolar interaction, $d_{ij}$, between the $M_i$ and $M_j$ nanomagnets were $d_{12}$ = 32.8 nm, $d_{23}$ = 47.7 nm and $d_{13}$ = 74.6 nm. The demagnetizing factors $N_{\sigma i}$ ($\sigma$ = x, y or z) were calculated by using approximate expressions for cylindrical magnets[S1]. The procedure of the estimation of $4\pi M_s$ (= 11.2 kOe) for $Co_{85}Cu_{15}$ is described in the main text.

Thus, the effective dipolar interactions being determined, we can consider an ideal system of three point dipoles. In such a system, there is no hysteresis. Therefore, the ferromagnetic resonance (FMR) frequencies as a function of applied magnetic field is symmetrical with respect to the $H$ = 0 axis (contrary to a real system). In the polar system of the coordinates, the position of each magnetic moment is defined by the polar angle $\theta$ and the azimuth $\varphi$ (the $\theta$ = 0 direction is along the nanowire axis). In our ideal system, all magnetic moments lie in one plane. As a consequence, the polar angle $\theta$ along is enough to describe the orientation of each magnet. To determine the FMR frequencies of individual magnets, we used the LLG equation of motion for a single-domain magnet in the small-precession-angle limit, expressed in terms of partial derivatives of the total free energy[S2,S3]:

$$f = \frac{\mu_B g}{h} \times \frac{1}{M_s \sin\theta_0} (F_{\theta\theta} F_{\varphi\varphi} - F_{\theta\varphi}^2)^{1/2},$$

where the second partial derivatives of the free energy, $F_{\theta\theta}$, $F_{\varphi\varphi}$ and $F_{\theta\varphi}$, are taken at the equilibrium positions of the magnetic moments $(\theta_0, \varphi_0)$. $\mu_B$ is the Bohr magneton, and $h$ is the Planck constant. In our case, $M_s$ is $M_{CoCu}$. We use the g-factor of pure cobalt $g$ = 2.25 (ref. S4).

As mentioned in the main text, in zero applied magnetic field, the measured FMR frequency of $M_2$ slightly varies: $5 \leq f_{FMR} \leq 7$ GHz. Numerical calculations (see below) show that $f_{FMR}$ (or $f_2$) is enough sensitive to the diameter of nanowires with a tendency: the larger the diameter, the higher the resonance frequency. At ø36 nm, $f_2$ = 5.6 GHz, while at ø40 nm, $f_2$ is already $f_2$ = 6.7 GHz. To fit the data in Figure 3 of the main text, in our numerical calculations the diameter of the nanowire was taken 36 nm. Also, independently of the diameter of the nanowires, the thickness of $M_1$ in calculations was reduced to 20 nm. This allowed a larger overlap of the regions of coexistence of FMR modes of the system <u>without</u> any significant effect on the FMR frequencies.

We found two static configurations which provide a complete description of the system. They are sketched at $H$ = 0 in Figure S3. The configuration 1 was found in a local minimum, while the configuration 2 in the global minimum, shown correspondently in Figures. S3a and S3b. The configuration 1 with *nearly* parallel orientations of the magnetic moments $M_1$ and $M_2$ is stable at high $|H|$ applied perpendicular to the nanowire axis. This configuration has a low resistance. The configuration 2 depicted in Figure S3 is stable at low $|H|$, and characterized by a higher resistance. In the configuration 2, the antiparallel orientation of the magnetic moments $M_1$ and $M_2$ is due to their dipolar interaction. The calculated FMR frequencies and corresponding angles for each configuration are shown in Figure S4. The FMR frequencies for each mode are symmetrical with respect to the $H = 0$ axis.

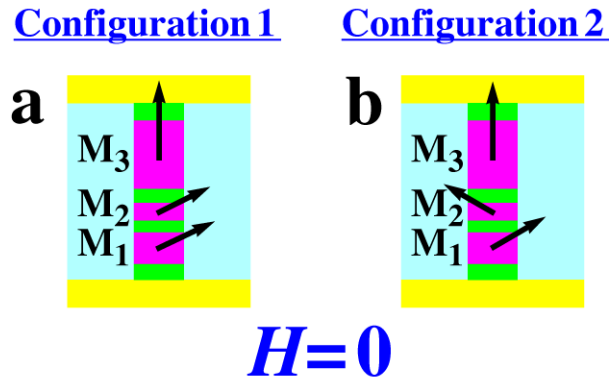

Figure S3. Two static stable configurations of a sandwiched PSV at $H = 0$, obtained by searching local minima of the free energy.

As shown in Figures. S4b and S4d, by applying a magnetic field perpendicular to the nanowire axis, the $M_3$ cylinder in both configurations undergoes a continuous rotation from $\theta = 90°$ to $\theta = -90°$ in the field changing from $(N_x - N_z)M_S$ to $-(N_x - N_z)M_S$ irrespectively of the orientation of the smaller magnets $M_1$ and $M_2$ (to recall, $N_x - N_z = 0.29$ [in units of $4\pi$]; see main text). According to the simulations, at small $|H|$, the $M_1$ and $M_2$ magnets are around $\pm 75°$ with respect to the nanowire axis. The dependence of the FMR frequency of $M_3$ on the applied field does not differ much from that of an isolated $M_3$ cylinder. At $H = 0$, we find $f_3 = 10.47$ GHz in the antiparallel orientation of $M_1$ and $M_2$ (measured $f_3$ is 10.4 GHz). The FMR frequencies of the smaller magnets at $H = 0$ are $f_2 = 5.61$ GHz and $f_1 = 1.83$ GHz (the measured $f_2$ in Fig.3a is 5.36 GHz). The relatively high $f_2$ frequency is due to the strong dipolar interaction with both $M_1$ and $M_3$ magnets.

In order to fit experimental data, namely, the magnetoresistance in Figure 3e, and the FMR frequencies in Figure 4a, the switching between the two configurations is done "manually". From the slope of the $f_2(H)$ dependence shown in Figure 4a (red curve), we conclude that the $M_1$ magnet undergoes reorientation first while sweeping the field (see Figure 3e). The high-frequency mode at weak negative fields in Figure 4a occurs due to simultaneous reorientation of both $M_1$ and $M_2$ magnets. The result of such an adjustment is shown in Figure 4c. To recall, our simulations deal exclusively with *uniform* FMR modes.

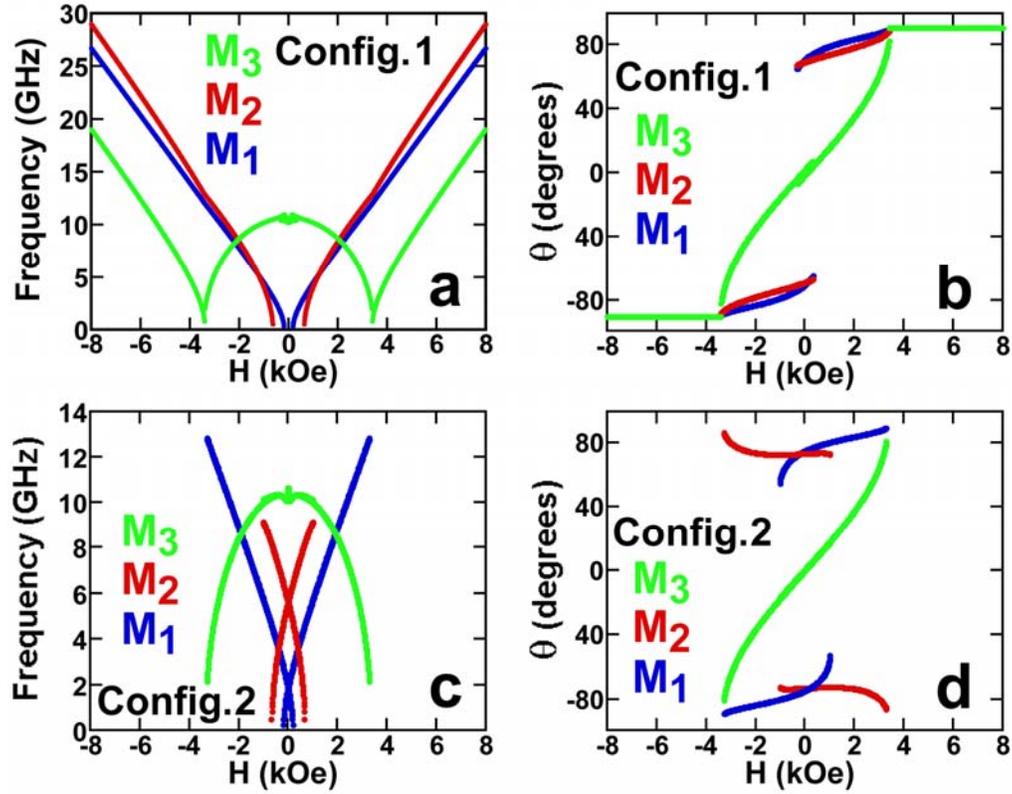

Figure S4. Numerical LLG computations of FMR frequencies (plots **(a)** and **(c)**) and angles between the magnetic moments $M_i$ ($i$ = 1,2 and 3) and the nanowire axis (plots **(b)** and **(d)**) for an *ideal* sandwiched PSV sketched in Figure 1b. The two configurations are shown in Figure S3. The color of the curves corresponds to the color of $M_i$ in the plots.

## 3. Complementary Measurements

In addition to the ac-current excited FMR measurements (spin-torque diode effect), we have also carried out measurements of dc-current excited steady state oscillations (SSO) of the free (middle) layer in the sandwiched pseudo-spin-valves (PSV). To detect a microwave signal of SSO, we utilized the standard heterodyne technique. We used the resolution bandwidth of 1.9 MHz. The amplified signal was measured either by a microwave diode or by a Hewlett-Packard microwave power meter. They yield identical results.

As an example, Figure S5a shows two microwave-power spectra obtained in a sandwiched PSV in zero applied magnetic field. In the sandwiched PSV, the SSO occur in *both* directions of the dc current. This is *an advantage* for potential applications. The FWHM of the curves in Figure S5a is about 30 MHz which is the narrowest width observed in our measurements. The measurements have been done at $I_{dc}$ = –0.85 mA (red) and $I_{dc}$ = 0.85 mA (blue); thus, above the critical values of the dc current which are $I_c^-$ = –0.72 mA and $I_c^+$ = 0.77 mA, correspondently. The values have been deduced from a measurement of the differential resistance of the sandwiched PSV as a function of dc current, d$V$/d$I$($I_{dc}$), depicted in Figure S5b. The value of $I_c$ = 0.75 mA is typical for our sandwiched PSV. Along with the main jumps in d$V$/d$I$($I_{dc}$), we often observed smaller peaks at lower values of the currents (–0.54 mA and 0.58 mA in Figure S5b). The origin of the smaller, inner peaks is unknown. The value of 0.85 mA corresponds in our nanostructures to the current density of about $6.5 \times 10^7$ A/cm$^2$.

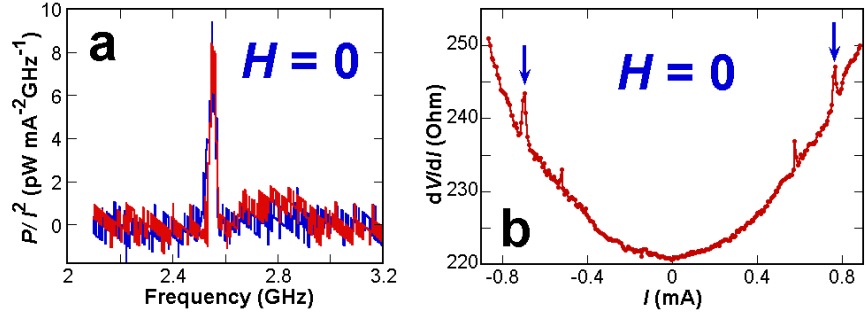

Figure S5. **(a)** Spectra of SSO obtained in a sandwiched PSV in zero applied magnetic field. Microwave power density curves are obtained by applying a dc current to a sandwiched PSV: $I_{dc}$ = –0.85 mA (red) and $I_{dc}$ = 0.85 mA (blue). The positive direction of the current corresponds to the current flow in the direction from the $M_1$ magnet to $M_3$ (see Figure 1a in main text). The resolution bandwidth is 1.9 MHz. The FWHM of the spectra is about 30 MHz. **(b)** Differential resistance of the same sandwiched PSV as a function of dc current in zero applied magnetic field. The arrows indicate the critical value $I_c$ for each direction of the current, $I_c^-$ = –0.72 mA and $I_c^+$ = 0.77 mA.